\documentclass[reprint,preprintnumbers,tightenlines,aip,apl]{revtex4-2}
\usepackage{graphicx}
\usepackage{adjustbox}
\usepackage{multirow}
\usepackage{latexsym,amssymb}
\usepackage{amsmath,bbold}
\usepackage{paralist}
\usepackage{braket}
\usepackage{xcolor}
\UseRawInputEncoding

\newcommand{\tphase}{P4$_{2}$nmc }
\newcommand{\oI}{Pca2$_{1}$ }
\newcommand{\oII}{Pmn2$_{1}$ }
\newcommand{\iCM}{cm$^{-1}$~}

\begin{document}

\title{First-principles Calculations of Raman and Infrared Spectroscopy For Phase Identification and Strain Calibration of Hafnia.}

\author{Aldo Raeliarijaona}
\email[Author to whom correspondence should be addressed: Aldo Raeliarijaona, ]{araeliarijaona@carnegiescience.edu}
\author{R. E. Cohen}
\email{rcohen@carnegiescience.edu}
\affiliation{Extreme Materials Initiative, Earth and Planets Laboratory, Carnegie Institution for Science, 5241 Broad Branch Road NW, Washington, DC 20015, USA}

\date{\today}

\begin{abstract}
Using density functional perturbation theory (DFPT) we computed the phonon frequencies, Raman and IR activities of hafnia polymorphs (\tphase, \oI, \oII, Pbca OI, brookite, and baddeleyite) for phase identification.
We investigated the evolution of Raman and IR activities with respect to epitaxial strain and provide plots of frequency differences as a function of strain for experimental calibration and identification of the strain state of the sample. We found Raman signatures of different hafnia polymorphs: $\omega(A_{1g})=300$ \iCM for \tphase, $\omega(A_{1})=343$ \iCM for \oI, $\omega(B_{2})=693$ \iCM for \oII, $\omega(A_{g})=513$ \iCM for Pbca (OI), $\omega(A_{g})=384$ \iCM for brookite, and $\omega (A_{g}) = 496$ \iCM for baddeleyite. We also identified the Raman $B_{1g}$ mode, an anti-phase vibration of dipole moments, ( $\omega(B_{1g}) = 758$ \iCM for OI, $\omega (B_{1g})= 784$ \iCM for brookite) as the Raman signature of antipolar Pbca structures. We calculated a large splitting between longitudinal optical (LO) and transverse optical (TO) modes ($\Delta{\omega_{\text{LO-TO}}(A^{z}_{1})}=255$ \iCM in \oI, and $\Delta{\omega_{\text{LO-TO}}(A_{1})}=263$ \iCM in \oII) to the same order as those observed in perovskite ferroelectrics, and related them to the anomalously large Born effective charges of Hf atoms ($Z^{*}(\text{Hf}) = 5.54$).
\end{abstract}
                          
\maketitle

Hafnia is of great interest because of its compatibility with silicon and its robust ferroelectricity at the nanoscale. It has a high permittivity and has been proposed as a dielectric material for dynamic random access memories (DRAM) and logic devices\cite{Kingon2000,Wilk2001}. At ambient pressure and temperature it stabilizes in the monoclinic crystal structure (P2$_{1}$/c) \cite{Ruh1970}, but can also have different structures depending on temperature and pressure. At ambient pressure there are two high-temperature polymorphs: the tetragonal (\tphase), when 1700$^{\circ}$C $<$ T $<$ 2600$^{\circ}$C, and the cubic ($Fm\bar{3}m$) when T $> $2600$^{\circ}$C\cite{Ohtaka2001}. Other phases have been stabilized under different pressure, such as Pbca, when 5.9 GPa $<$ P $<$ 15 GPa, and Pnma when under P $>$ 15 GPa\cite{Ohtaka2001,Huan2014}, epitaxial strain\cite{Liu2019}, or doping\cite{Park2017,Batra2017}.

Recent investigations have highlighted the possibility and importance of ferroelectricity in hafnia\cite{Boscke2011,schroeder_FE2019,Lee2020,Qi2020,Xu2021,Mikolajick2021}, and studies of yttrium-doped hafnia revealed the \oI polymorph as the polar phase responsible for ferroelectricity\cite{Muller2011}. Antiferroelectricity has also been observed in doped hafnia \cite{ParkAFE2015} and further theoretical studies have shown that the orthorhombic Pbca structure, which contains two \oI cells with opposite oxygen displacement relative to the hafnia atoms, as the antiferroelectric orthorhombic structure\cite{Lee2020}.
Hafnia is also of fundamental interest due to its fluorite-based structure, different from the wide class of well-studied perovskite ferroelectrics. One hindrance to the rapid development of hafnia ferroelectrics has been a lack of basic understanding of ferroelectricity.

Structural studies are usually based on X-ray diffraction (XRD) methods. Complementary to XRD, inelastic light scattering such as Raman or IR spectroscopy is also a tool that can be used to determine the structure and structural response of different hafnia phases. The latter approach reveals information about interatomic interaction, hence atomic arrangements, through the vibrational spectra. Epitaxial strain is important in thin films and it can affect the Raman frequencies and intensities. One of the advantages of Raman spectroscopy is that it can provide information about the strain on the sample once calibrated. Only Raman shifts and peak intensities are accessible experimentally thus the need of first-principles studies of atomic vibrations for symmetry assignment of Raman peaks. Although Raman and IR data are available from previous calculations of hafnia polymorphs\cite{Zhou2014,Gunst2020,Fan2022} they do not cover all the relevant phases, namely both polar orthorhombic phases or both antipolar orthorhombic phases. Furthermore, experimentally it is normal to observe mixture of different phases in hafnia such as baddeleyite and \oI \cite{Mukundan2021}, which makes the symmetry determination using Raman or IR spectroscopy crucial.

Our aim in this paper is to provide tools for experiments to identify the structure of hafnia and determine the epitaxial strain to which the sample is subject. Using DFPT we computed the phonon frequencies and their activities (Raman and IR) at different strain values ($\eta = [-3.0\%, -1.5\%, 0.0\%, 1.5\%, 3.0\%]$) for different hafnia polymorphs, namely baddeleyite, \tphase, \oII, \oI, Pbca (OI), and brookite. We tracked the evolution of phonon mode frequencies under strain, and provided means to identify the strain through identification of strain-induced frequency shifts of signature phonons.

We performed first-principles calculations using {\sc{Quantum Espresso}} \cite{QE-2009,QE-2017,QE-exa}, with optimized norm-conserving Vanderbilt (ONCV) pseudopotentials \cite{Hamann2013}. The cell parameters and atomic positions were optimized using the local density approximation (LDA) with the Perdew-Zunger (PZ) exchange-correlation functional \cite{PZ1981}, generated using the {\sc oncvpsp} code \cite{Hamann2013}. The plane-wave expansion is truncated using a cutoff energy of E$_\textbf{cutoff}$ =  1306.6 eV, and the Brillouin zone was sampled using an 8$\times$8$\times$8 Monkhorst-Pack grid \cite{MP1976}. We calculated the Brillouin zone center phonon frequencies using density functional perturbation theory (DFPT) \cite{Baroni2001} implemented in the QE/PH package \cite{QE-2009,QE-2017,QE-exa}. We considered the powder Raman spectra, with the Raman activity computed using Placzek approximation\cite{Porezag1996,Lazzeri2003,Caracas2006}. We plotted the spectra by convoluting with a Gaussian shape function whose width is 5 \iCM estimated from experiments. Similar widths have been used in previous studies\cite{Caracas2006}.

We set a coordinate system similar to the ones defined in Ref.\onlinecite{Liu2019}. We consider the epitaxy of the tetragonal case with a square substrate such as Yttria-stabilized zirconia (YSZ), with \textbf{z} normal to the epitaxial plane, and \textbf{a} and \textbf{b} in registry with the epitaxial surface. Strain $\eta$ was applied by setting:
\begin{equation}\label{EqStrain}
a_{\eta} =  (1+\eta) a_{0},
\end{equation}
where $a_{0}$ is the strain-free ground state lattice constant of the tetragonal phase. The out-of-plane lattice constant ($c$) and the atomic positions were relaxed at each strain.

\begin{figure}[h!]
\centering
\includegraphics[width=0.85\columnwidth]{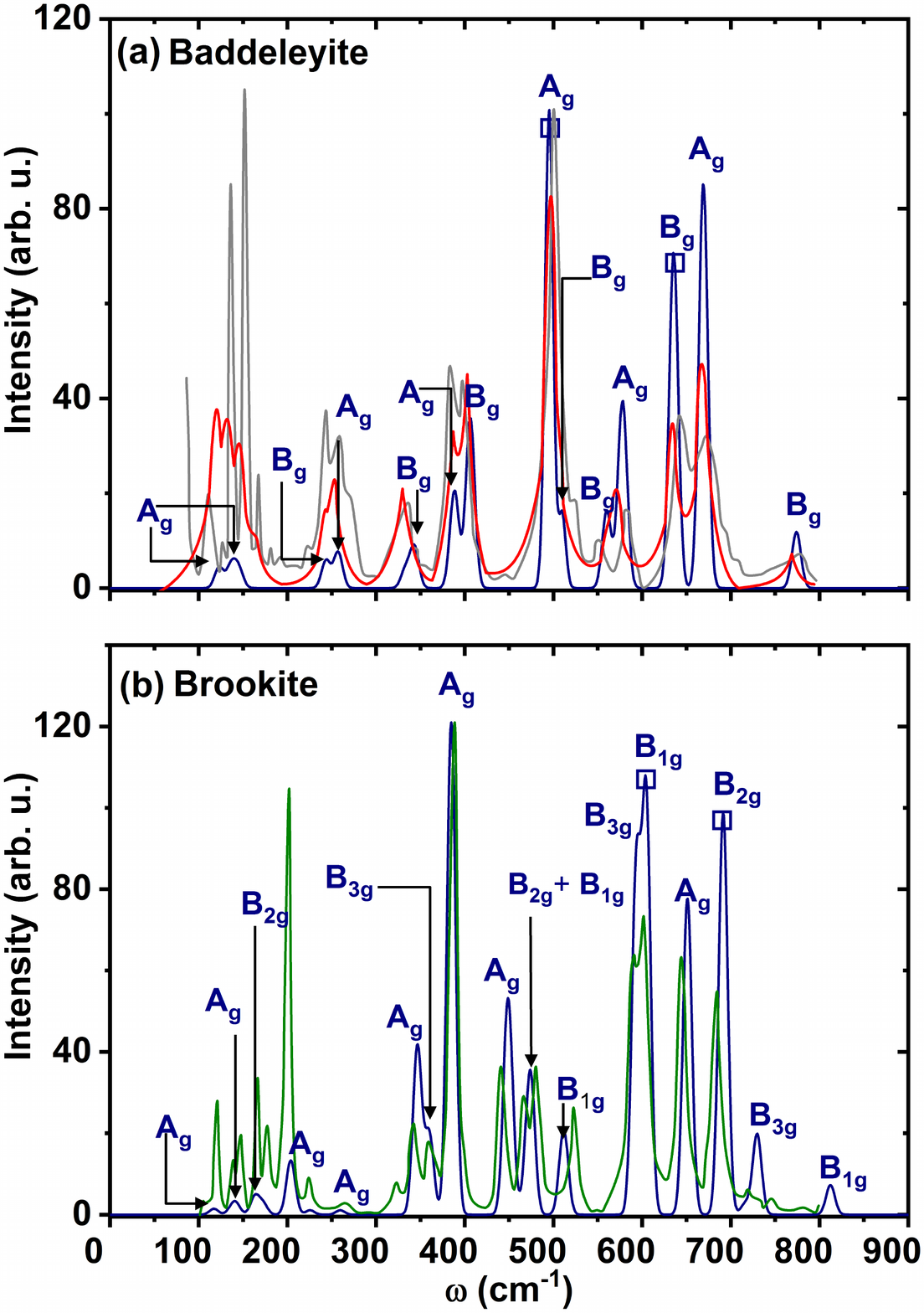}
\caption{Raman spectra of \textbf{(a)} baddeleyite,\textbf{(b)} brookite compared to experimental and theoretical Raman. Gray and green spectra are experimental data from Ref.\onlinecite{Jayaraman1993} and the red curve is a theoretical spectrum from Ref.\onlinecite{Zhou2014}. We convoluted the spectra with a Gaussian shape function of width 5 \iCM estimated from experiments. The squares indicate modes chosen to compute the frequency differences (Fig.\ref{Fig4}). For brookite Raman spectra \textbf{(b)}, the blue and green curves were obtained under hydrostatic pressure (P = 5.9 GPa).}
\label{Fig1}
\end{figure}

To validate the accuracy of our DFPT calculations, we compare the calculated Raman spectra of the strain-free baddeleyite and brookite to the measured Raman spectra\cite{Jayaraman1993} (Fig.\ref{Fig1}). The Raman spectrum of brookite\cite{Jayaraman1993} was measured from a sample under hydrostatic pressure of P = 5.9 GPa. So, for fair comparison, the cell parameters and the atomic positions for our calculations for brookite were relaxed under the same pressure. The calculated spectra of baddeleyite and brookite match the experimentally measured spectra well (Fig.\ref{Fig1}). For baddeleyite, the high intensity $A_{g}$ peaks ($\omega(A_{g}) = 495$ \iCM ), the shoulder peak $B_{g}$ ($\omega(B_{g}) = 510$ \iCM ), and the double peaks $B_{g}$ ($\omega(B_{g}) =637$ \iCM)  and $A_{g}$ ($\omega(A_{g}) = 670$ \iCM) all closely match with experimental peaks; for brookite we can point to the high intensity $A_{g}$ peak ($\omega(A_{g}) = 384$ \iCM),  the $B_{1g}$ peak ($\omega(B_{1g}) = 513$ \iCM), the double peaks $B_{3g}$ ($\omega(B_{3g}) = 593$ \iCM) and $B_{1g}$ ($\omega(B_{g}) = 604$ \iCM). Our calculated spectra for baddeleyite, tetragonal and brookite also compare well with other calculations\cite{Zhou2014,Gunst2020} (see Supplementary Material Fig.S1).

\begin{figure}[h!]
\centering
\includegraphics[width=1.0\columnwidth]{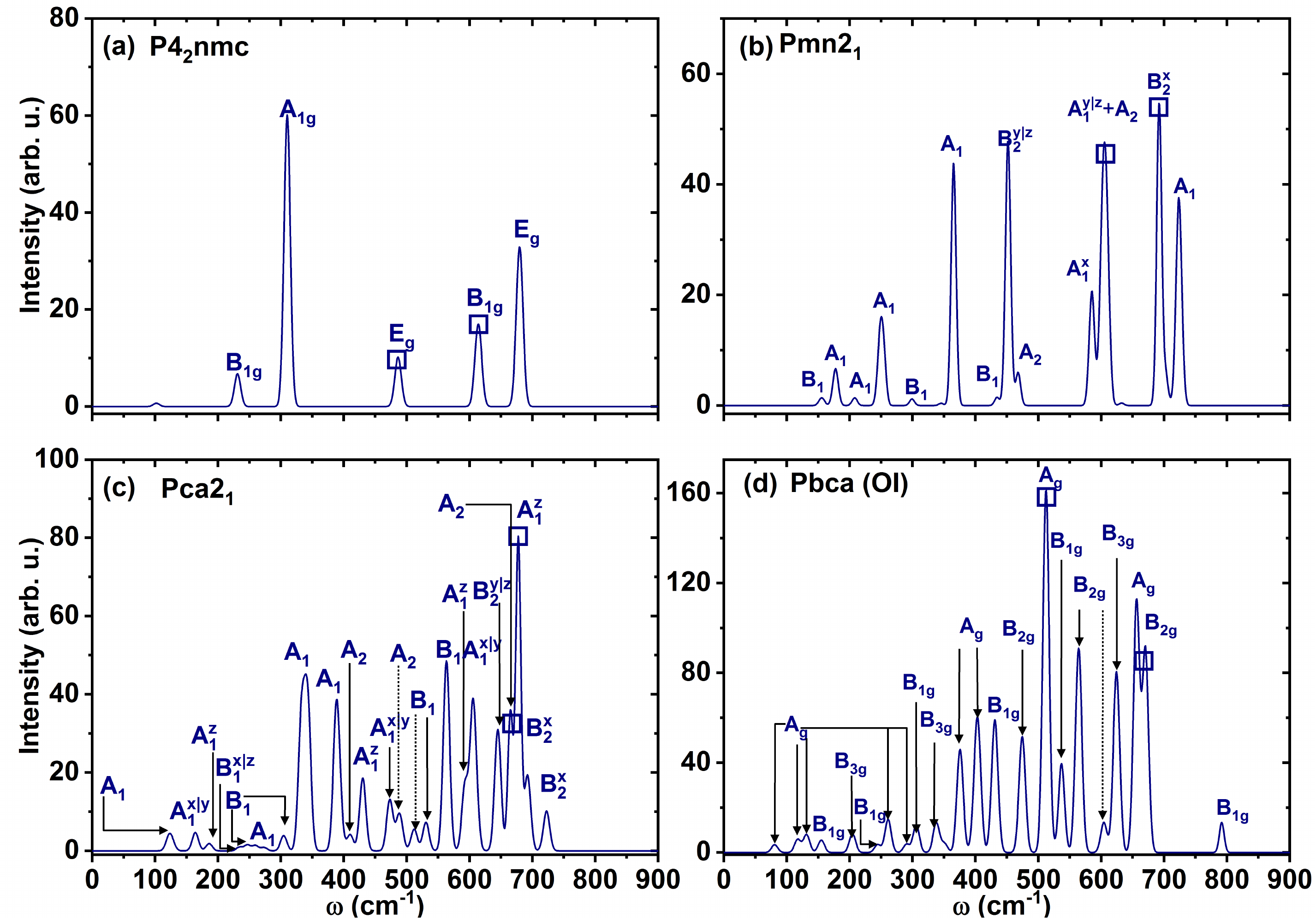}
\caption{Raman spectra of \textbf{(a)} \tphase, \textbf{(b)} \oII, \textbf{(c)} \oI, and \textbf{(d)} Pbca (OI). We convoluted the spectra using a Gaussian shape function of width 5 \iCM estimated from experiments. The squares indicate modes chosen to compute the frequency differences (Fig.\ref{Fig4}).}
\label{Fig2}
\end{figure}

We plot the Raman spectra of the different strain-free hafnia polymorphs (Fig.\ref{Fig1},Fig.\ref{Fig2}) to help with the identification of the crystal structures. The high intensity peaks of the Raman spectra can help in the identification of the structure symmetry. For example, the high intensity mode at $\omega(A_{g}) = 496$ \iCM is found only in baddeleyite. The \oII structure can be identified using the high frequency $B_{2}$ ($\omega(B_{2}) = 693$ \iCM) and $A_{1}$ ($\omega(A_{1}) = 722$ \iCM) peaks. The \oI structure can be identified using the $A_{1}$ peak ($\omega = 340$ \iCM) and the strong $A_{1}$ peak ($\omega = 670$ \iCM). The presence of these peaks can be used to identify the \oI structure. This is an advantage of the Raman (or IR) spectroscopy over the XRD given that phase identification using XRD can be harder due to the monoclinic and \oI phase mixtures and the overlap of XRD peaks \cite{Mukundan2021}, whereas with Raman spectroscopy the difference can be resolved easily. In Pbca (OI) the signature modes are the strong $A_{g}$ peaks at $\omega(A_{g})=513$ \iCM, and the double $A_{g}$, and $B_{2g}$ peaks at $\omega(A_{g})=658$ \iCM and $\omega(B_{2g})=668$ \iCM, respectively. Finally, in brookite the $A_{g}$ mode ($\omega(A_{g}) = 384$ \iCM)  can be used as its fingerprint.

\begin{figure}[h!]
\centering
\includegraphics[width=0.65\columnwidth]{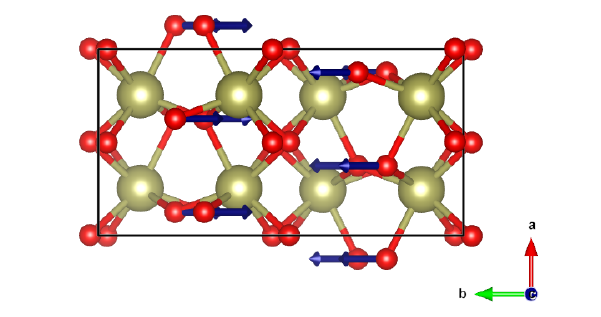}
\caption{The atomic displacements corresponding to the high frequency $B_{1g}$ mode in Pbca AFE structures. The larger spheres represent the Hf atoms and the smaller spheres the O atoms. The blue arrows on the atoms indicate the direction of the atomic vibrations. The arrows on the bottom right corner show the crystallographic directions.}
\label{Fig3}
\end{figure}

Additionally, we find a signature mode for antipolar polymorphs. This is the high frequency B$_{1g}$ peak ($\omega(B_{1g})= 811$ \iCM for brookite, or $\omega(B_{1g})=791$ \iCM for Pbca OI). This mode is absent in the FE (\oI and \oII) structures and it is an anti-phase vibration of the O atoms in neighboring cells of the Pbca (Fig.\ref{Fig3}). This anti-phase vibration of the O atoms can be thought as an anti-phase collective motion of the dipole moments.

The splitting between longitudinal optical (LO) and transverse optical (TO) phonon modes is a subject of interest in ferroelectrics. Indeed, perovskite ferroelectrics are known to have phonon modes that exhibit giant LO-TO splittings\cite{Zhong1994}. Using the rigorous definition of LO-TO splitting\cite{Fu2015}, we show that these large LO-TO splittings also occur in hafnia. In polar \oII phase, the calculated LO-TO splitting is $\Delta{\omega_{\text{LO-TO}}}(A	_{1})=263$ \iCM for the high frequency $A_{1}$ mode ($\omega(A_{1})=722$ \iCM); for the \oI phase we computed the LO-TO splitting to be $\Delta{\omega_{\text{LO-TO}}(A^{z}_{1})}=255$ \iCM for $\omega(A^{z}_{1})=676$ \iCM and $\Delta{\omega_{\text{LO-TO}}(B^{x}_{2})}=146$ \iCM for the $B^{x}_{2}$ mode at $\omega(B^{x}_{2})=722$ \iCM (Fig.\ref{Fig2}). These large LO-TO splittings can be attributed to the unusually large Born effective charge ($Z^{*}$) of Hf atoms in hafnia (see Supplementary Material). Although the born effective charges $Z^{*}(O)$ are close to the nominal charge of O atoms (-2), with few exceptions such as: $Z^{*}_{yy}(O1)=Z^{*}_{xx}(O2)=-3.42$ in \tphase, $Z^{*}_{xx}(O1)=-3.15$ and $Z^{*}_{zz}(O2)=-3.08$ in \oII, the Hf atoms have anomalously large dynamical charges up to 5.54. Other calculations also reported similar values\cite{Neal2021,Fan2022}. The three modes with large LO-TO splitting mentioned earlier all have in common the antiphase motion between the Hf and O.

The constraint on the lattice constant imposed by the substrate affects the interatomic distances in the sample, which not only shifts the phonon frequencies but also changes the Raman and IR intensities. The shift in Raman or IR frequencies thus indicates the strains in a hafnia film. The evolution of phonon modes with respect to strain can be tracked using correspondence between the phonon eigenvectors at each strain following the approach in Ref.\onlinecite{Fu2015}. Firstly, the phonon eigenvectors at strain $\eta_{2}$ were projected to the eigenvectors at strain $\eta_{1}$:
\begin{equation}\label{ModeCorres}
\epsilon_{n}^{i\alpha}(\eta_{2})=\sum\limits_{m} a_{mn} \epsilon_{m}^{i\alpha}(\eta_{1}) ,
\end{equation}
where $m$ and $n$ are the phonon mode indices, $i$ and $\alpha$ are the atoms and the direction indices respectively; the coefficient $a_{mn} = \sum\limits_{i\alpha}\langle\epsilon_{m}^{i\alpha}(\eta_{1}) \vert \epsilon_{n}^{i\alpha}
(\eta_{2}) \rangle$ indicates the correlation or projection of phonon eigenmode $n$ at strain $\eta_{2}$ to the eigenmodes at strain $\eta_{1}$. Then by choosing the mode with the maximum coefficient $a_{mn}$ (Eq.~\ref{ModeCorres}), i.e. the mode with the highest correspondence probability to mode $n$ denoted by $\omega^{m^{\prime}}_{\eta1}$, we assign $\omega^{n}_{\eta2} \rightarrow  \omega^{m^{\prime}}_{\eta1}$.

The calculation of the frequency difference is straightforward once equipped with the strain-to-strain mode correspondence (Eq.\ref{ModeCorres}):
\begin{equation}\label{Calibration}
\Delta{\omega_{l,n}(\eta)}= \omega_{l^{\prime}}(\eta) - \omega_{n^{\prime}}(\eta),
\end{equation}
where it should be understood that $\omega_{l^{\prime}}(\eta)$, and $\omega_{n^{\prime}}(\eta)$ are the phonon modes that correspond to $\omega_{l}(0)$, and $\omega_{n^{\prime}}(0)$ at zero strain, respectively.
$\Delta{\omega_{l,n}}(\eta)$ changing sign means that the order of the modes considered switched.

\begin{figure}[!ht]
\centering
\includegraphics[width=\columnwidth]{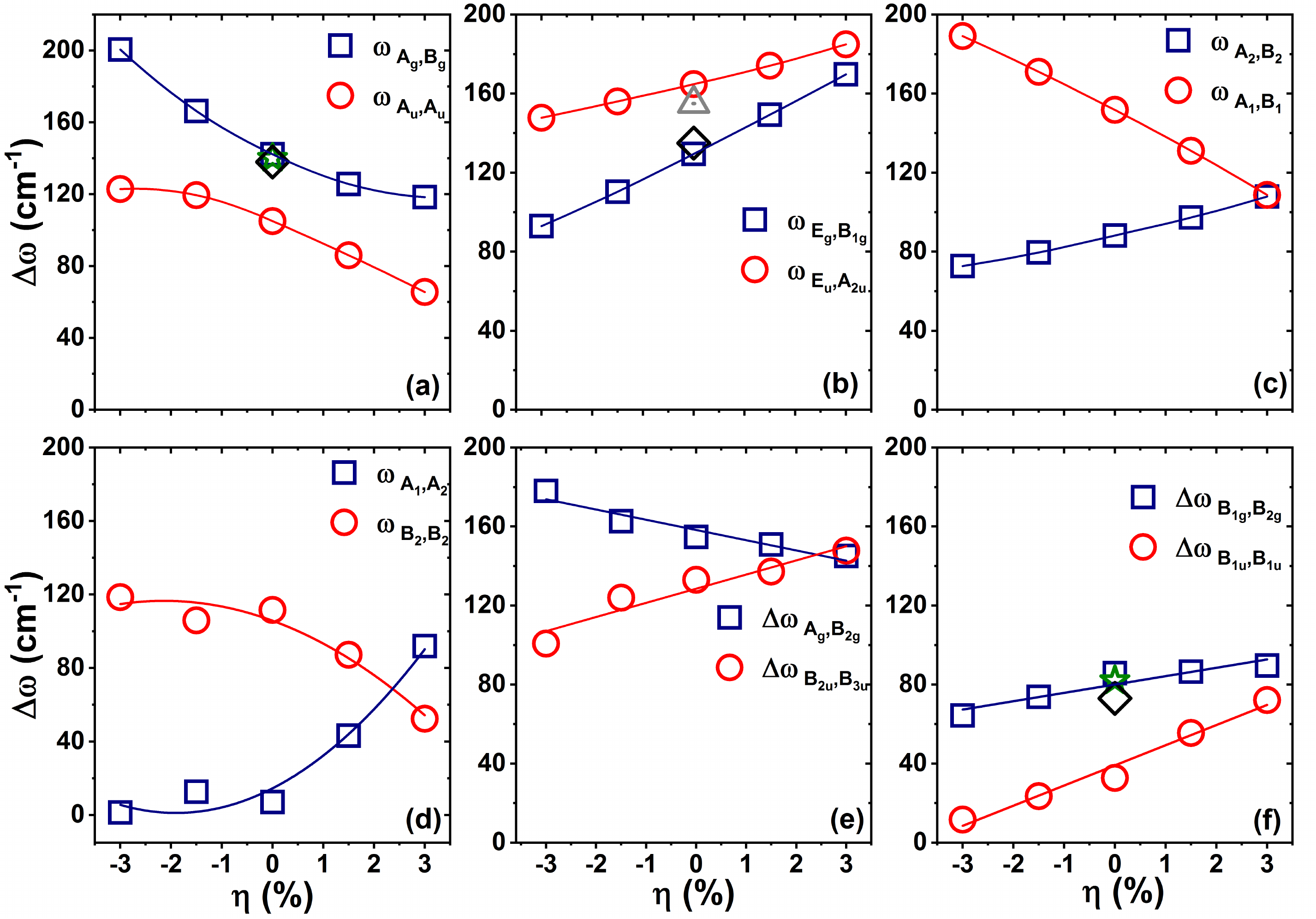}
\caption{Evolution of frequency difference with respect to epitaxial strain for selected modes of \textbf{(a)} baddeleyite, \textbf{(b)} \tphase, \textbf{(c)} \oII, \textbf{(d)} \oI, \textbf{(e)} Pbca OI, and \textbf{f} brookite. The star, diamond, and triangle symbols in \textbf{(a)}, \textbf{(b)}, and  \textbf{(f)} are frequency differences using data from Ref.\onlinecite{Jayaraman1993}, Ref.\onlinecite{Zhou2014}, and Ref.\onlinecite{Yashima1996}, respectively. The blue and red curves are calibration using IR and Raman active modes, respectively. The Raman modes used for calibrations are marked by the empty squares on the Raman spectra in Fig\ref{Fig1} and the IR modes are marked as blue modes in Table S3--S8 (see Supplementary Material).}
\label{Fig4}
\end{figure}

The evolution of the phonon modes is not straightforward, and choices had to be made on the modes used to compute the frequency difference (Fig.\ref{Fig4}). The modes chosen (Fig.~\ref{Fig1}, and Fig.\ref{Fig2}) were modes with non-negligible Raman or IR intensities and the ones showing monotonic behavior if possible, within a range of 200 \iCM for plotting purposes. In Fig.\ref{Fig4} we used results from studies with samples that are either under hydrostatic pressure\cite{Jayaraman1993} or doped hafnia\cite{Yashima1996} for comparison. For the case where doping helps with the stabilization of a particular polymorph, peaks associated to defects or dopants were identified\cite{Yashima1996} so the comparison with the remaining peaks, probably from the tetragonal polymorph, with our calculation results can be made.

For the comparison of Pbca brookite spectra, the Raman spectra in Ref.\onlinecite{Jayaraman1993} were obtained under hydrostatic pressure. We performed two different first-principles calculations for the brookite phase, one at ambient pressure and one under 5.9 GPa. We find that pressure hardens the individual Raman modes, as expected. But the difference between two modes ($\Delta{\omega_{l,n}}$) for the system at ambient pressure was similar to $\Delta{\omega_{l,n}}$ for the system under hydrostatic pressure. It is thus fair to compare $\Delta{\omega_{l,n}}$ from the experiment under pressure with $\Delta{\omega_{l,n}}$ from DFPT at ambient pressure (Fig.\ref{Fig4}f).

For most cases, the frequency difference between two peaks changes linearly with strain, such as $\Delta{\omega_{B_{1g},B_{2g}}}(\eta) = 80 +4.2 \eta $ for the Raman active modes chosen in brookite or $\Delta{\omega_{A_{2},B_{2}}}(\eta) = 89 +5.9 \eta $ for the Raman active modes chosen in \oII. 
On the other hand, for the \oI and baddeleyite the frequency difference is better described by a quadratic polynomial of the strain $\eta$:
\begin{equation}
\Delta{\omega_{A_{g},B_{g}}}(\eta) = 142 - 13.7 \eta + 2 \eta^{2},
\end{equation}
for the Raman active modes chosen in baddeleyite, and
\begin{equation}
\Delta{\omega_{A_{1},B_{1}}}(\eta) = 15 +14.1 \eta + 3.7 \eta^{2} 
\end{equation}
for \oI.

Now we briefly discuss an interesting comparison between hafnia and zirconia, two isomorphous compounds\cite{Ruh1970}. Recent studies on ZrO$_{2}$ combined experimental measurements and computations of Raman spectra to identify the zirconia polymorphs using Raman\cite{Materano2022}. The Raman spectra of monoclinic hafnia and zirconia are similar, e.g.\ the $A_{g}$ peak at $\omega(A_{g})=495$ \iCM in hafnia and the $A_{g}$ peak at $\omega(A_{g})=470$ \iCM in zirconia, or the $B_{g}$ peak at $\omega(B_{g})=637$ \iCM in hafnia compared to $\omega(B_{g})=598$ \iCM in zirconia. The quantitative mismatch between the spectra in hafnia and zirconia (25 \iCM for $A_{g}$ and 30 \iCM  for B$_{g}$) can be attributed to differences in the interatomic force constant (IFC) resulting from the difference in the cation. Interestingly, despite these quantitative discrepancies in the Raman frequencies of hafnia and zirconia, the frequency differences $\Delta{\omega_{l,n}}$ are comparable. In monoclinic hafnia for example, the frequency difference $\Delta{\omega_{A_{g},B_{g}}}=142$ \iCM compares well with $\Delta{\omega_{A_{g},B_{g}}}=140$ \iCM in zirconia, and in tetragonal the frequency difference $\Delta{\omega_{E_{g},B_{1g}}}=130$ \iCM in hafnia is a close match to $\Delta{\omega_{E_{g},B_{1g}}}=135$ \iCM in zirconia.

We computed the phonon frequencies of \tphase, \oII, \oI, Pbca (OI), brookite, and baddeleyite using DFPT at different values of epitaxial strains. We found that Raman spectra can be used to identify the symmetry or phases of hafnia, namely $\omega(A_{1g})=300$ \iCM for \tphase, $\omega(A_{1})=343$ \iCM for \oI, $\omega(B_{2})=693$ \iCM for \oII, $\omega(A_{g})=513$ \iCM for Pbca (OI), $\omega(A_{g})=384$ \iCM for brookite, and $\omega (A_{g}) = 496$ \iCM for baddeleyite. We also identified the Raman signature of AFE structures as the $\omega(B_{1g}) =784$ \iCM peak for brookite or the $\omega(B_{1g}) =758$ \iCM peak for Pbca (OI). The presence (or absence) of this Raman signal can be used to distinguish between AFE and FE orthorhombic hafnia. 
Further, we showed the evolution of frequency differences between selected normal mode frequencies with respect to strain for calibration purposes and identification of strain state of the hafnia crystal.

\section*{Supplementary Material}
Comparison of bookite Raman spectra at ambient pressure(Fig.S1), the calculated Born effective charges (Table S1--S2), the infrared frequencies and IR activities of different hafnia polymorphs (Table S3--S8).

\begin{acknowledgments}
The authors thank Pavlo Zubko for helpful discussions.
 This work is supported by U. S. Office of Naval Research Grants No. N00014-17-1-2768 and N00014-20-1-2699, and the Carnegie Institution for Science. Computations were supported by DOD HPC, Carnegie computational resources, and REC gratefully acknowledges the Gauss Centre for Supercomputing e.V. (www.gauss-center.eu) for funding this project by providing computing time on the GCS Supercomputer SuperMUC-NG at Leibniz Supercomputing Centre (LRZ, www.lrz.de).
Some of the computing for this project was performed on the Memex cluster. We would like to thank Carnegie Institution for Science and the Carnegie Sci-Comp Committee for providing computational resources and support that contributed to these research results.
\end{acknowledgments}
\bibliography{Hafnia}
\end{document}


                          
\section*{\NoCaseChange{Supplementary Material: First-principles Calculations of Raman and Infrared Spectroscopy For Phase
Identification and Strain Calibration of Hafnia.}}

\subsection{Comparison of Raman spectra}

\begin{figure}[h!]
\centering
\includegraphics[width=0.85\columnwidth]{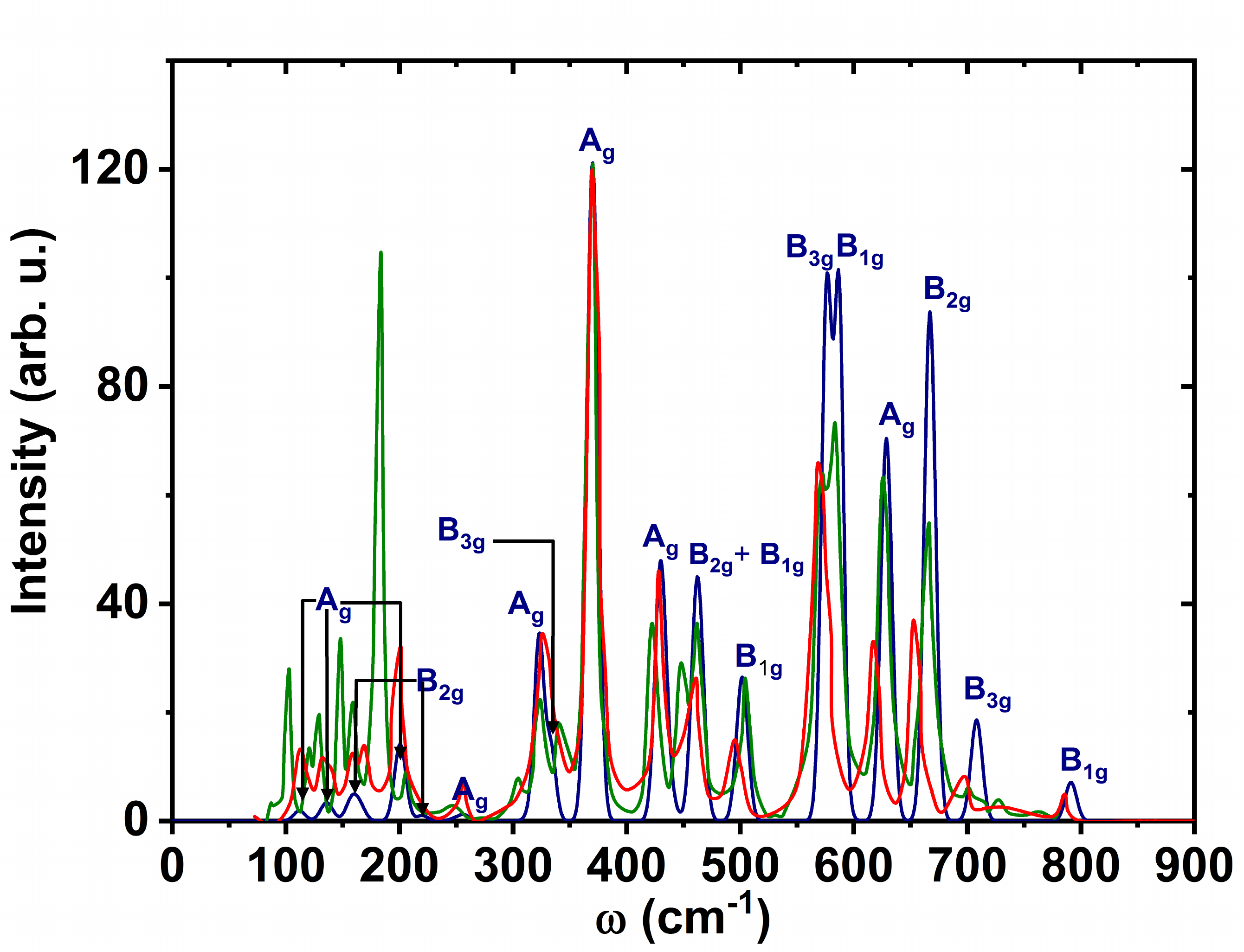}
\caption{Calculated Raman spectrum of brookite (blue) compared to theoretical Raman spectrum (red)\cite{Zhou2014S} at ambient pressure. The calculated spectrum was convoluted with a Gaussian shape function of width 5 \iCM estimated from experiments.}
\label{FigS1}
\end{figure}

\newpage
\subsection{Born effective charges}

\begin{center}
\begin{table}[!ht]\label{BornEffC_1}
\begin{tabular}{|c|c|c|c|c|c|c|c|c|c|c|c|}
\hline
Phase & Atoms  & $Z^{*}_{xx}$  & $Z^{*}_{xy}$ & $Z^{*}_{xz}$ &  $Z^{*}_{yx}$ & $Z^{*}_{yy}$ & $Z^{*}_{yz}$ &  $Z^{*}_{zx}$&  $Z^{*}_{zy}$ & $Z^{*}_{zz}$  & Nominal charge \\
\hline
\multirow{6}{*}{Baddeleyite} & Hf1 & 5.33 &-0.39 & 0.23 & -0.12 & 5.34 & 0.15  & 0.24 & 0.35 & 4.75 & 4\\
& Hf2 & 5.33 & 0.39 & 0.23 & 0.12 & 5.34 & -0.15  & 0.24 & -0.35  & 4.75 & 4 \\
& O1  & -2.95 & -1.05 & -0.22 & -1.30 & -2.60 & 0.64 & -0.19 & 0.63 & -2.22 & -2 \\
& O2  & -2.95 & 1.05 & -0.22 & 1.30 & -2.60 & -0.64 & -0.19 & -0.63  & -2.22 & -2   \\
& O3  & -2.37 & -0.11 & -0.01 & -0.18 & -2.74 & -0.35 & -0.05 & -0.42 & -2.53 & -2 \\
& O4  & -2.37 & 0.11 & -0.01 & 0.18  & -2.74 & 0.35 & -0.05 & 0.42  & -2.53 & -2  \\
\hline
\multirow{3}{*}{\tphase} & Hf & 5.54 & 0 & 0 & 0 &  5.54 & 0 & 0 & 0  & 4.84 & 4\\
& O1   & -2.12 & 0 & 0 & 0 & -3.42 & 0 & 0 & 0  & -2.42 & -2 \\
& O2   & -3.42 & 0 & 0 & 0 & -2.12 & 0 & 0 & 0  & -2.42 & -2\\
\hline
\multirow{6}{*}{\oII} & Hf1 & 5.35 & 0 & 0 & 0 &  5.02 & 0.10 & 0 & 0.33 & 5.11 & 4 \\
& Hf2 & 5.35 & 0 & 0 & 0 &  5.02  & -0.10 & 0 & -0.33 & 5.11 & 4 \\
& O1  & -3.15 & 0 & 0 & 0 & -2.67 & -0.80 & 0 & -0.86  & -2.03 & -2 \\
& O2  & -3.15 & 0 & 0 & 0 & -2.67 & 0.80 & 0 & 0.86 & -2.03 & -2\\
& O3  & -2.20 & 0 & 0 & 0 & -2.35 & 0.04 & 0 & -0.05  & -3.08 & -2\\
& O4  & -2.20 & 0 & 0 & 0 & -2.35 & -0.04 & 0 &  0.05 & -3.08 & -2\\
\hline
\end{tabular}
\caption{Computed Born effective charges for baddeleyite, \tphase and \oII hafnia polymorphs compared to their nominal charges. Hf1, Hf2 (resp. O1, O2 , O3, and O4 ) are the symmetrically inequivalent Hf (resp. O) atoms in the HfO$_{2}$ unit cell.} 
\end{table}
\end{center}

\begin{center}
\begin{table}[!ht]\label{BornEffC_2}
\begin{tabular}{|c|c|c|c|c|c|c|c|c|c|c|c|}
\hline
Phase & Atoms  & $Z^{*}_{xx}$  & $Z^{*}_{xy}$ & $Z^{*}_{xz}$ &  $Z^{*}_{yx}$ & $Z^{*}_{yy}$ & $Z^{*}_{yz}$ &  $Z^{*}_{zx}$&  $Z^{*}_{zy}$ & $Z^{*}_{zz}$  & Nominal charge \\
\hline
\multirow{12}{*}{\oI} & Hf1 & 5.17 & 0.00 & -0.035 & -0.35 &  5.49 & -0.14 & 0.05 & -0.22  & 4.97 & 4\\
& Hf2 & 5.17 & 0.00 & 0.035 & -0.35 &  5.49 & 0.14 & -0.05 & 0.22 & 4.97 & 4\\
& Hf3 & 5.17 & 0.00 & -0.035 & 0.35 &  5.49 & 0.14 & 0.05 & 0.22 & 4.97 & 4\\
& Hf4 & 5.17  & 0.00 & 0.035 & 0.35  &  5.49 & -0.14 & -0.05 & -0.22 & 4.97 & 4\\
& O1  & -2.49 & -0.96 & 0.31 & -0.76 & -2.96 & -0.69 & 0.31 & -0.63  & -2.46 & -2\\
& O2  & -2.49 & -0.96 & -0.31 & -0.76 & -2.96 & 0.69 & -0.31 & 0.63 & -2.46 & -2 \\
& O3  & -2.49 & 0.96 & 0.31 & 0.76 & -2.96 & 0.69 & 0.31 & 0.63  & -2.46 & -2\\
& O4  & -2.49 & 0.96 & -0.31 & 0.76 & -2.96 & -0.69 & -0.31 & -0.63 & -2.46 & -2\\
& O5  & -2.68 & 0.13 & 0.25 & 0.07 & -2.52 & -0.19 & 0.29 & -0.13 & -2.50 & -2 \\
& O6  & -2.68 & 0.13 & -0.25 & 0.07 & -2.52 & 0.19 & -0.29 & 0.13 & -2.50 & -2  \\
& O7  & -2.68 & -0.13 & 0.25 & -0.07 & -2.52 & 0.19 & 0.29 & 0.13 & -2.50 & -2  \\
& O8  & -2.68 & -0.13 & -0.25 & -0.07 & -2.52 & -0.19 & -0.29 & -0.13 & -2.50 & -2  \\
\hline
\multirow{12}{*}{Pbca (OI)} & Hf1 & 5.28 & 0.34 & 0.13 & 0.34 &  4.78  & -0.03 & 0.30 & -0.08 & 5.43 & 4 \\
& Hf2 & 5.28 & 0.34 & -0.13 & 0.34 &  4.78 & 0.03 & -0.30 & 0.08 & 5.43 & 4 \\
& Hf3 & 5.28 & -0.34 & 0.13 &-0.34 &  4.78 & 0.03 & 0.30 & 0.08 & 5.43 & 4 \\
& Hf4 & 5.28 & -0.34 & -0.13 & -0.34 &  4.78 & -0.03 & -0.30 & -0.08 & 5.43 & 4 \\
& O1  & -2.56 & -0.87 & -1.14 & -0.87 & -2.23 & 0.01 & -0.89 & 0.10  & -3.02 & -2\\
& O2  & -2.56 & -0.87 & 1.14 & -0.87 & -2.23 & -0.01 & 0.89 & -0.10 & -3.02 & -2 \\
& O3  & -2.56 & 0.87 & -1.14 & 0.87 & -2.23 & -0.01 & -0.89 & -0.10 & -3.02 & -2 \\
& O4  & -2.56 & 0.87 & 1.14 & 0.87 & -2.23 & 0.01 & 0.89 & 0.10  & -3.02 & -2\\
& O5  & -2.73 & 0.28 & -0.12 & 0.41 & -2.55 & -0.10 & -0.07 & -0.02  & -2.41 & -2 \\
& O6  & -2.73 & 0.28 & 0.12 & 0.41 & -2.55 & 0.10 & 0.07 & 0.02  & -2.41 & -2\\
& O7  & -2.73 & -0.28 & -0.12 & -0.41 & -2.55 & 0.10 & -0.07 & 0.02  & -2.41 & -2\\
& O8  & -2.73 & -0.28 & 0.12 & -0.41 & -2.55 & -0.10 & 0.07 & -0.02  & -2.41 & -2\\
\hline
\multirow{12}{*}{Brookite} & Hf1 & 4.89 & 0.24 & -0.02 & 0.04 &  5.46 & 0.39 & -0.13 & -0.00 & 5.31 & 4 \\
& Hf2 & 4.89 & 0.24 & 0.02 & 0.04 &  5.46 & -0.39 & 0.13 & 0.00 & 5.31 & 4\\
& Hf3 & 4.89 & -0.24 & -0.02 & -0.04 &  5.46 & -0.39 & -0.13 & 0.00 & 5.31 & 4\\
& Hf4 & 4.89 & -0.24 & 0.02 & -0.04 &  5.46 & 0.39 & 0.13 & -0.00 & 5.31 & 4 \\
& O1  & -2.49 & -0.04 & 0.36 & -0.09 & -2.49 & 0.04 & 0.31 & 0.01 & -2.76 & -2 \\
& O2  & -2.49 & -0.04 & -0.36 & -0.09 & -2.49 & -0.04 & -0.31 & -0.01 & -2.76 & -2 \\
& O3  & -2.49 & 0.04 & 0.36 & 0.09 & -2.49 & -0.04 & 0.31 & -0.01  & -2.76 & -2 \\
& O4  & -2.49 & 0.04 & -0.36 & 0.09 & -2.49 & 0.04 & -0.31 & 0.01  & -2.76 & -2 \\
& O5  & -2.39 & -0.56 & 0.31 & -0.62 & -2.97 & -0.80 & 0.32 & -0.98 & -2.55 & -2 \\
& O6  & -2.39 & -0.56 & -0.31 & -0.62 & -2.97 & 0.80 & -0.32 & 0.98 & -2.55 & -2  \\
& O7  & -2.39 & 0.56 & 0.31 & 0.62 & -2.97 & 0.80 & 0.32 & 0.98  & -2.55 & -2 \\
& O8  & -2.39 & 0.56 & -0.31 & 0.62 & -2.97 & -0.80 & -0.32 & -0.98 & -2.55 & -2 \\
\hline
\end{tabular}
\caption{Computed Born effective charges for \oI, Pbca (OI) and brookite hafnia polymorphs compared to their nominal charges. Hf1, Hf2, Hf3 and Hf4 (resp. O1, O2 , O3, and O4 ) are the symmetrically inequivalent Hf (resp. O) atoms in the HfO$_{2}$ unit cell.} 
\end{table}
\end{center}

\subsection{IR frequencies}
Here we list the computed IR frequencies, their intensities and their symmetry.

\begin{center}
\begin{table}[h!]
\centering
\begin{adjustbox}{width=\columnwidth,center}
\begin{tabular}{|c|c|c|c|c|c|c|c|c|c|c|c|}
\hline
 & Mode & Freq (cm$^{-1}$) & IR (arb. u.) &  & Mode & Freq (cm$^{-1}$) & IR (arb. u.) &  & Mode & Freq (cm$^{-1}$) & IR (arb. u.) \\
\hline
\multirow{12}{*}{q $\rightarrow$[001]} & B$_{u}$ &  236  & 22.84 & \multirow{12}{*}{q $\rightarrow$[010]} & B$_{u}$ &  232  & 20.07 & \multirow{12}{*}{q $\rightarrow$[100]}  & B$_{u}$ &  249  &  5.72 \\
& \R{A$_{u}$} &  \R{256}  & \R{20.18} & & B$_{u}$ &  311  & 64.83 &  & A$_{u}$ &  256  & 20.18 \\
& B$_{u}$ &  320  & 41.87 & & B$_{u}$ &  345  & 24.74 &  & B$_{u}$ &  256  &  2.02 \\
& \R{A$_{u}$} &  \R{361}  & \R{39.70} & & B$_{u}$ &  405  & 40.35 &  & B$_{u}$ &  345  & 23.81 \\
& B$_{u}$ &  370  &  7.43 & & A$_{u}$ &  476  &  2.40 &  & A$_{u}$ &  361  & 39.70 \\
& A$_{u}$ &  411  & 24.29 & & B$_{u}$ &  520  & 19.66 &  & B$_{u}$ &  400  & 49.06 \\
& A$_{u}$ &  499  & 12.55 & & A$_{u}$ &  575  &  9.21 &  & A$_{u}$ &  411  & 24.29 \\
& B$_{u}$ &  516  & 19.40 & & B$_{u}$ &  741  & 14.87 &  & B$_{u}$ &  455  &  9.53 \\
& A$_{u}$ &  614  & 12.53 & & A$_{u}$ &  750  & 96.89 &  & A$_{u}$ &  499  & 12.55 \\
& B$_{u}$ &  651  & 76.23 & &         &       &       &  & A$_{u}$ &  614  & 12.53 \\
& B$_{u}$ &  743  & 16.88 & &         &       &       &  & B$_{u}$ &  623  & 34.20 \\
&         &       &       & &         &       &       &  & B$_{u}$ &  813  & 60.68 \\
\hline
\end{tabular}
\end{adjustbox}
\caption{Frequencies of IR-active modes in baddeleyite. Modes colored in blue are the modes used to compute frequency differences (see text).}
\label{STable1}
\end{table}
\end{center}


\begin{center}
\begin{table}[h!]
\centering
\begin{adjustbox}{width=\columnwidth,center}
\begin{tabular}{|c|c|c|c|c|c|c|c|c|c|c|c|}
\hline
 & Mode & Freq (cm$^{-1}$) & IR (arb. u.) &  &  Mode & Freq (cm$^{-1}$) & IR (arb. u.) & & Mode & Freq (cm$^{-1}$) & IR (arb. u.) \\
\hline
\multirow{5}{*}{q $\rightarrow$[001]} &  E$_{u}$ &  95   & 19.57 & \multirow{5}{*}{q $\rightarrow$[010]} &  E$_{u}$ &  95   & 19.57 & \multirow{5}{*}{q $\rightarrow$[100]}  &  E$_{u}$ &  95   & 19.57 \\
&  E$_{u}$ &  95   & 19.57 & &  E$_{u}$ &  246  & 2.62 &  &  E$_{u}$ &  246  & 2.62 \\
&  E$_{u}$ &  462  & 35.04 & & A$_{2u}$ &  324  & 39.86 & & A$_{2u}$ &  324  & 39.86 \\
& \R{E$_{u}$} &  \R{462}  &\R{35.04} & &  E$_{u}$ &  462  & 35.04 & &  E$_{u}$ &  462  & 35.04 \\
& \R{A$_{2u}$} & \R{627}  &\R{39.86} & &  E$_{u}$ &  720  & 51.99 & &  E$_{u}$ &  720  & 51.99 \\
\hline
\end{tabular}
\end{adjustbox}
\caption{Frequencies of IR-active modes for \tphase. Modes colored in blue are the modes used to compute  frequency differences (see text).}
\label{STable2}
\end{table}
\end{center}

\begin{center}
\begin{table}[h!]
\centering
\begin{adjustbox}{width=\columnwidth,center}
\begin{tabular}{|c|c|c|c|c|c|c|c|c|c|c|c|}
\hline
& Mode & Freq (cm$^{-1}$) & IR (arb. u.) & & Mode & Freq (cm$^{-1}$) & IR (arb. u.) & & Mode & Freq (cm$^{-1}$) & IR (arb. u.) \\
\hline
\multirow{10}{*}{q $\rightarrow$[001]} & B$_{2}$ & 250 & 25.10 & \multirow{10}{*}{q $\rightarrow$[010]}  & A$_{1}$ & 177 &  6.80 & \multirow{10}{*}{q $\rightarrow$[100]}  & A$_{1}$ & 177 &  6.80 \\
& B$_{1}$ & 300 & 32.60 & & B$_{2}$ & 250 & 25.10 & & A$_{1}$ & 254 &  3.80 \\
& B$_{2}$ & 451 & 24.87 & & A$_{1}$ & 254 &  3.80 & & \R{B$_{1}$} & \R{300} & \R{32.60} \\
& B$_{1}$ & 492 & 10.10 & & A$_{1}$ & 364 &  2.93 & & B$_{2}$ & 345 &  2.01 \\
& A$_{1}$ & 587 &  3.01 & & B$_{1}$ & 432 &  3.06 & & A$_{1}$ & 364 &  2.93 \\
& B$_{1}$ & 701 &  2.10 & & A$_{1}$ & 451 & 31.75 & & \R{A$_{1}$} & \R{451} & \R{31.75} \\
& A$_{1}$ & 722 & 43.82 & & B$_{2}$ & 451 & 24.87 & & B$_{1}$ & 492 & 10.10 \\
&         &     &       & & A$_{1}$ & 606 &  2.84 & & A$_{1}$ & 606 &  2.84 \\
&         &     &       & & B$_{1}$ & 632 & 20.16 & & B$_{2}$ & 692 & 47.97 \\
&         &     &       & & B$_{1}$ & 738 & 20.90 & & B$_{1}$ & 701 &  2.10 \\
\hline
\end{tabular}
\end{adjustbox}
\caption{Frequencies of IR-active modes for \oII. Modes colored in blue are the modes used to compute frequency differences (see text).}
\label{STable3}
\end{table}
\end{center}

\begin{center}
\begin{table}[h!]
\centering
\begin{adjustbox}{width=\columnwidth,center}
\begin{tabular}{|c|c|c|c|c|c|c|c|c|c|c|c|}
\hline
& Mode & Freq (cm$^{-1}$) & IR (arb. u.) &   &  Mode & Freq (cm$^{-1}$) & IR (arb. u.) &  & Mode & Freq (cm$^{-1}$) & IR (arb. u.) \\
\hline
\multirow{16}{*}{q $\rightarrow$[001]}  &  B$_{1}$ &  144  &  1.06 & \multirow{16}{*}{q $\rightarrow$[010]} &  A$_{1}$ &  122  &  2.18 &  \multirow{16}{*}{q $\rightarrow$[100]}  &  A$_{1}$ &  122  &  2.18 \\
&  B$_{1}$ &  237  & 40.77 & &  A$_{1}$ &  163  &  8.47 & &  B$_{1}$ &  144  &  1.06 \\
&  B$_{1}$ &  248  &  1.29 & &  A$_{1}$ &  260  &  2.19 & &  A$_{1}$ &  163  &  8.47 \\
&  B$_{2}$ &  275  & 37.59 & &  B$_{2}$ &  275  & 37.59 & &  B$_{1}$ &  237  & 40.77 \\
&  B$_{2}$ &  335  &  2.14 & &  B$_{1}$ &  305  &  2.09 & &  B$_{1}$ &  248  &  1.29 \\
&  B$_{2}$ &  393  & 50.69 & &  A$_{1}$ &  333  &  5.98 & &  A$_{1}$ &  260  &  2.19 \\
&  B$_{1}$ &  397  & 49.25 & &  B$_{2}$ &  335  &  2.14 & &  A$_{1}$ &  333  &  5.98 \\
&  A$_{1}$ &  431  &  3.16 & &  A$_{1}$ &  342  & 46.83 & &  A$_{1}$ &  342  & 46.83 \\
&  B$_{2}$ &  490  &  1.94 & &  B$_{2}$ &  393  & 50.69 & &  B$_{1}$ &  397  & 49.25 \\
&  B$_{1}$ &  532  &  8.19 & &  A$_{1}$ &  471  & 19.92 & &  A$_{1}$ &  471  & 19.92 \\
&  A$_{1}$ &  593  &  8.10 & &  B$_{2}$ &  490  &  1.94 & &  B$_{1}$ &  532  &  8.19 \\
&  B$_{2}$ &  644  &  3.79 & &  B$_{1}$ &  510  &  4.39 & &  A$_{1}$ &  606  &  1.95 \\
&  A$_{1}$ &  676  & 75.48 & &  A$_{1}$ &  606  &  1.95 & &  \R{B$_{2}$} &  \R{610}  & \R{22.60} \\
&  B$_{1}$ &  747  &  7.78 & &  B$_{2}$ &  644  &  3.79 & &  B$_{2}$ &  692  & 22.93 \\
&          &       &       & &  B$_{1}$ &  652  & 49.35 & &  \R{B$_{2}$} &  \R{722}  & \R{50.20} \\
&          &       &       & &  B$_{1}$ &  796  & 53.18 & &  B$_{1}$ &  747  &  7.78 \\
\hline
\end{tabular}
\end{adjustbox}
\caption{Frequencies of IR-active modes for \oI. Modes colored in blue are the modes used to compute frequency differences (see text).}
\label{STable4}
\end{table}
\end{center}

\begin{center}
\begin{table}[h!]
\centering
\begin{adjustbox}{width=\columnwidth,center}
\begin{tabular}{|c|c|c|c|c|c|c|c|c|c|c|c|}
\hline
 & Mode & Freq (cm$^{-1}$) & IR (arb. u.) & & Mode & Freq (cm$^{-1}$) & IR (arb. u.) & & Mode & Freq (cm$^{-1}$) & IR (arb. u.) \\
\hline
\multirow{18}{*}{q $\rightarrow$[001]} & B$_{3u}$ &  249  &  15.35 & \multirow{18}{*}{q $\rightarrow$[010]} & B$_{1u}$ &  233  &  15.82 & \multirow{18}{*}{q $\rightarrow$[100]}  & B$_{1u}$ &  233  &  15.82 \\
& B$_{2u}$ &  266  &  43.84 & & B$_{1u}$ &  262  &   9.72 & & B$_{3u}$ &  249  &  15.35 \\
& B$_{3u}$ &  313  &   9.06 & & B$_{2u}$ &  266  &  43.84 & & B$_{1u}$ &  262  &   9.72 \\
& B$_{2u}$ &  323  &  56.19 & & B$_{1u}$ &  296  & 115.67 & & B$_{1u}$ &  296  & 115.67 \\
& \R{$_{3u}$} &  \R{379}  & \R{127.96} & & B$_{2u}$ &  323  &  56.19 & & B$_{3u}$ &  313  &   9.06 \\
& B$_{2u}$ &  386  &  64.30 & & B$_{1u}$ &  358  &  20.51 & & B$_{2u}$ &  353  &   1.43 \\
& B$_{3u}$ &  454  &   4.70 & & B$_{2u}$ &  386  &  64.30 & & B$_{1u}$ &  358  &  20.51 \\
& B$_{1u}$ &  471  &  16.78 & & B$_{2u}$ &  512  &  30.16 & & B$_{3u}$ &  379  & 127.96 \\
& B$_{3u}$ &  477  &   7.51 & & B$_{1u}$ &  515  &  22.85 & & B$_{3u}$ &  454  &   4.70 \\
& \R{B$_{2u}$} &  \R{512}  &  \R{30.16} & & B$_{1u}$ &  580  &   2.29 & & B$_{2u}$ &  476  &   8.89 \\
& B$_{1u}$ &  573  &   7.54 & & B$_{2u}$ &  617  &  19.59 & & B$_{3u}$ &  477  &   7.51 \\
& B$_{2u}$ &  617  &  19.59 & & B$_{3u}$ &  653  & 154.48 & & B$_{1u}$ &  515  &  22.85 \\
& B$_{1u}$ &  623  &  73.73 & & B$_{1u}$ &  742  &  25.26 & & B$_{1u}$ &  580  &   2.29 \\
& B$_{1u}$ &  805  & 113.23 & & B$_{3u}$ &  811  &   9.98 & & B$_{2u}$ &  584  &  17.46 \\
& B$_{3u}$ &  806  &   2.00 & &          &       &        & & B$_{2u}$ &  669  &   2.11 \\
&          &       &        & &          &       &        & & B$_{1u}$ &  742  &  25.26 \\
&          &       &        & &          &       &        & & B$_{2u}$ &  742  & 184.42 \\
&          &       &        & &          &       &        & & B$_{3u}$ &  806  &   2.00 \\
\hline
\end{tabular}
\end{adjustbox}
\caption{Frequencies of IR-active modes in Pbca OI. Modes colored in blue are the modes used to compute frequency differences (see text).}
\label{STable5}
\end{table}
\end{center}

\begin{center}
\begin{table}[h!]
\centering
\begin{adjustbox}{width=\columnwidth,center}
\begin{tabular}{|c|c|c|c|c|c|c|c|c|c|c|c|}
\hline
\multirow{19}{*}{q $\rightarrow$[001]} & Mode & Freq (cm$^{-1}$) & IR (arb. u.) & \multirow{17}{*}{q $\rightarrow$[010]} & Mode & Freq (cm$^{-1}$) & IR (arb. u.) & \multirow{16}{*}{q $\rightarrow$[100]} &  Mode & Freq (cm$^{-1}$) & IR (arb. u.)  \\
\hline
& B$_{2u}$ &  191  &  17.37 & & B$_{2u}$ &  191  &  17.37 & & B$_{3u}$ &  253  &   8.16 \\
& B$_{2u}$ &  224  &   5.85 & & B$_{2u}$ &  224  &   5.85 & & B$_{3u}$ &  263  &  79.64 \\
& B$_{3u}$ &  253  &   8.16 & & B$_{1u}$ &  280  &  71.48 & & B$_{1u}$ &  280  &  71.48 \\
& B$_{3u}$ &  263  &  79.64 & & B$_{3u}$ &  319  &   2.92 & & B$_{3u}$ &  385  &  88.62 \\
& B$_{1u}$ &  323  &   2.28 & & B$_{2u}$ &  363  &  51.56 & & \R{B$_{1u}$} &  \R{392}  &  \R{82.96} \\
& B$_{2u}$ &  363  &  51.56 & & B$_{2u}$ &  368  &  55.47 & & B$_{2u}$ &  423  &   1.81 \\
& B$_{2u}$ &  368  &  55.47 & & B$_{1u}$ &  392  &  82.96 & & \R{B$_{1u}$} &  \R{425}  &  \R{33.01} \\
& B$_{3u}$ &  385  &  88.62 & & B$_{1u}$ &  425  &  33.01 & & B$_{1u}$ &  506  &   8.66 \\
& B$_{2u}$ &  441  &  30.44 & & B$_{2u}$ &  441  &  30.44 & & B$_{3u}$ &  555  &  21.39 \\
& B$_{1u}$ &  499  &   1.97 & & B$_{1u}$ &  506  &   8.66 & & B$_{2u}$ &  571  &  19.38 \\
& B$_{3u}$ &  555  &  21.39 & & B$_{3u}$ &  517  &  16.09 & & B$_{1u}$ &  646  &   7.35 \\
& B$_{2u}$ &  594  &   8.16 & & B$_{2u}$ &  594  &   8.16 & & B$_{2u}$ &  669  &  12.62 \\
& B$_{1u}$ &  620  &  31.91 & & B$_{1u}$ &  646  &   7.35 & & B$_{2u}$ &  683  & 133.57 \\
& B$_{1u}$ &  703  &  66.91 & & B$_{3u}$ &  658  &  85.45 & & B$_{1u}$ &  726  &   2.32 \\
& B$_{1u}$ &  750  & 102.39 & & B$_{1u}$ &  726  &   2.32 & & B$_{3u}$ &  752  &  17.74 \\
& B$_{3u}$ &  752  &  17.74 & & B$_{3u}$ &  808  & 111.77 & &          &       &        \\
\hline
\end{tabular}
\end{adjustbox}
\caption{Frequencies of IR-active modes in brookite. Modes colored in blue are the modes used to compute frequency differences (see text).}
\label{STable6}
\end{table}
\end{center}

\FloatBarrier
\bibliography{HafniaSup}